\documentclass[11pt,a4paper]{article}
\usepackage[]{amsmath,amssymb}
\usepackage{graphics,epsfig}
\begin{document}
\date{}
\title{Reduced density matrix and internal dynamics for multicomponent regions}
\author{H. Casini\footnote{e-mail: casini@cab.cnea.gov.ar}
 \, and M. Huerta\footnote{e-mail: marina.huerta@cab.cnea.gov.ar} \\
{\sl Centro At\'omico Bariloche,
8400-S.C. de Bariloche, R\'{\i}o Negro, Argentina}}
\maketitle

\begin{abstract}
We find the density matrix corresponding to the vacuum state of a massless Dirac field in two dimensions reduced to a region of the space formed by several disjoint intervals. We calculate explicitly its spectral decomposition. The imaginary power of the density matrix is a unitary operator implementing an internal time flow (the modular flow). We show that in the case of more than one interval this evolution is non-local, producing both, advance in the causal structure and "teleportation" between the disjoint intervals.  However, it only mixes the fields on a finite number of trajectories, one for each interval. As an application of these results we compute the entanglement entropy for the massive multi-interval case in the small mass limit.      
\end{abstract}

\section{Introduction}
An observer with constant self-acceleration $a$ in vacuum Minkowski space-time sees himself immersed in a thermal bath at temperature $T=a/(2\pi)$ \cite{unruh}. This is the Unruh effect. One explanation for this phenomenon is that the constant acceleration impedes this observer to be in contact with all the Minkowski space. In fact, there is a portion of the space-time (a Rindler Wedge), including half of a spatial hyperplane, from which the signals cannot reach the observer.  In consequence, the degrees of freedom lying in this region are not relevant for the description of the physics in the observer's world. If we trace over them, the vacuum state becomes a mixed state on the observer's half space $V$. This mixing reflects the presence of quantum entanglement between the seen and the hidden regions. 

In a more general case, we can consider localization in a different region $V$. The corresponding local state $\rho_V$ is defined by the property that it leads to the same expectation values for the operators localized in $V$ than the vacuum state. In presence of a regularization, $\rho_V$ can be written in terms of a density matrix, in such a way we have
\begin{equation}
\textrm{tr}( \rho_V {\cal O})=\langle 0|{\cal O}|0\rangle\,,\label{an}
\end{equation} 
for any operator ${\cal O}$ localized in the region $V$ (or more precisely localized in the domain of dependence of $V$). 
In this paper we will assume implicitly such a regularization (a lattice regularization for example), which will allow us to use the language of local density matrices (in the continuum limit the local algebras of operators are a special kind of von Neumann algebras, and the states on the local algebras should be treated with more sofisticated mathematical machinery than density matrices).
    
Any density matrix can be written as an exponential  
 \begin{equation}
\rho _{V}=c \,e^{-{\cal H}}\,,\label{rho1}
\end{equation}
of a Hermitian operator ${\cal H}$, and where $c$ is a normalization constant. The thermal interpretation of the local state $\rho_V$ arises when one considers an internal time $\tau$, with unitary evolution given by the operator $\rho_V^{i \tau}$, inside the algebra of the fields in the region $V$. With respect to this notion of time evolution, $\rho_V$ is indeed thermal, with temperature rescaled to 1.

In the algebraic axiomatic approach to quantum field theory, the internal time evolution is called the modular flow and the Hermitian operator  corresponding to ${\cal H}$ the modular Hamiltonian \cite{haag}. These are well defined using the Tomita-Takesaki theory \footnote{In the formalism of quantum fields, the Tomita-Takesaki theory can be interpreted in terms of thermofield dynamics \cite{kleban}.}, which plays an important role in the structure. The ciclic and separating vector state of the Tomita-Takesaki theory (here the vacuum state $\left| 0 \right>$) is related to $\rho_V$ through eq. (\ref{an}).

In a different context, this mechanism of time flow generation from a thermodynamical state has been related with the time problem in quantum gravity \cite{connes}. The thermal time hypotesis aims to explain the emergence of a physical time flow in a general covariant quantum theory as determined by the thermodynamical state rather than been a universal property of the mechanical theory.

In the Unruh effect the set $V$ corresponds to a Rindler wedge. This particular case has a remarkable feature. The Hamiltonian has the universal expression ${\cal H}=2\pi K$, in terms of the boost operator $K$ which keeps the Rindler wedge fixed. This is valid for any quantum field theory \cite{unruhotro}. The internal evolution is local and causal. It is given by point transformations along the integral curves of the boost operator, which are precisely the trajectories of constant acceleration observers. In curved space-time a similar universality occurs for the black hole, where the local hamiltonian for the outside region and the Hartle-Hawking state is proportional to the time translation operator for asymptotic observers.  

For different type of sets these features should be lost in general. An exception is the internal evolution corresponding to spherical sets $V$ (which are single intervals in two dimensions) in conformal theories. This is also local, and given by the one parameter group of global conformal transformations which keep $V$ fixed \cite{haag,confor}.  However, no other explicit example of internal evolution is known in quantum field theory. In particular no non local example has been explicitly computed.

In this work we provide one such example. We calculate $\rho_V$ when  $V$ is formed by several disjoint spatial intervals in two dimensional Minkowski space, for a massless Dirac field. We explicitly diagonalize the reduced density matrix in this case, and obtain the internal evolution. We find some surprising features. Even if the internal evolution is non local, it does not mix all the operators on the Cauchy surface, but the operators in selected points on each interval. Also, we find that the total posible teleportation between the different regions appears to be limited. 

In curved space a similar phenomena should be present in the case of several black hole regions. 
In this context, the present example provides an exactly solvable toy model in order to explore the subtle effects of the multipartite entanglement present on the vacuum fluctuations, and their relation with the geometry. Also, these effects should give place in flat space to correlations on the  meassurements of two or more accelerated observers.

As an application of the results, we also calculate the multicomponent entanglement entropy in the massive case under some approximations. We show the property of extensivity of the mutual information which holds for the massless case is lost in the massive one.       

\section{Structure of the density matrix}
An expression for the reduced density matrix $\rho _{V}$ in terms of the two point correlators is known for free Dirac fields \cite{pes,ara,otro}. This follows from the fact that the expectation
values of polynomials on the fields located in $V$ computed with the help of $\rho _{V}$ must obey Wick's theorem (here the global state is the vacuum). This fixes 
$\rho _{V}$ to be of the form (\ref{rho1}) 
where ${\cal H}$ is a Hamiltonian for $V$ which is quadratic on the fields inside this region. For discrete bosonic systems analogous expressions for ${\cal H}$ exist which are quadratic on the field and momentum operators. However, the generalization of these expressions to the continuum case is not known \cite{nunu}.    

In this paper we are interested in free Dirac fields in two dimensions. However, the formulas in this section apply to any dimensions with the obvious modifications. Let us consider the general case of regions which are arbitrary spatial curves, not necessarily included in a single spatial line, as shown in figure 1. Thus we write ${\cal H}$ as
\begin{equation}
{\cal H}=\int_V ds_1\,ds_2 \,\Psi_i^{\dagger}(s_1)H_{ij}(s_1,s_2) \Psi_j(s_2)\,\,\,\,;\,\,\,\,\,H^{*}_{ji}(s_2,s_1)=H_{ij}(s_1,s_2)\,,\label{rho2}
\end{equation}
where $\Psi(s)\equiv \Psi(x^\mu(s))$, and $s_1,s_2$ are length parameters along the spatial curve $V$. 
The Dirac field $\Psi(x)$ satisfies the Dirac equation
\begin{equation}
(i \gamma^\mu \partial_\mu  -m) \, \Psi =0\,,
\end{equation}
and the 
canonical (equal-time) anticommutation relations
\begin{equation}
\left. \left\{\Psi_i(x,t),\Psi^{\dagger}_j(y,t^\prime)\right\}\right|_{t=t^\prime}=\delta(x-y)\delta_{ij}\,.\label{hacha}
\end{equation}
For the fields $\Psi(s)$ parametrized with the length $s$ on the surface $V$ the slope of the curve (its relative Lorentz boost with respect to the $t=$constant lines) has to be taken into account, and equation (\ref{hacha}) generalizes to
\begin{equation} \left. \left\{\Psi_i(s_1),\tilde{\Psi}_j(s_2)\right\}\right|_V=\delta(s_1-s_2)\delta_{ij}\,,\label{algebra}
\end{equation}
where we have written
\begin{equation}
\tilde{\Psi}_j(s)=\Psi_j^\dagger(s) \bar{\eta}(s) \,,\hspace{1.2cm} \bar{\eta}(s)=\gamma^0 \gamma^\mu \eta_\mu(s)\,,
\end{equation}
and $\eta_\mu (s)$ the unit vector normal to the surface $V$ at the point $s$. Note that $\bar{\eta}(s)=\bar{\eta}(s)^\dagger$ is hermitian and $\bar{\eta}(s)^2=1$.

\begin{figure} [t]
\centering
\leavevmode
\epsfysize=5cm
\bigskip
\epsfbox{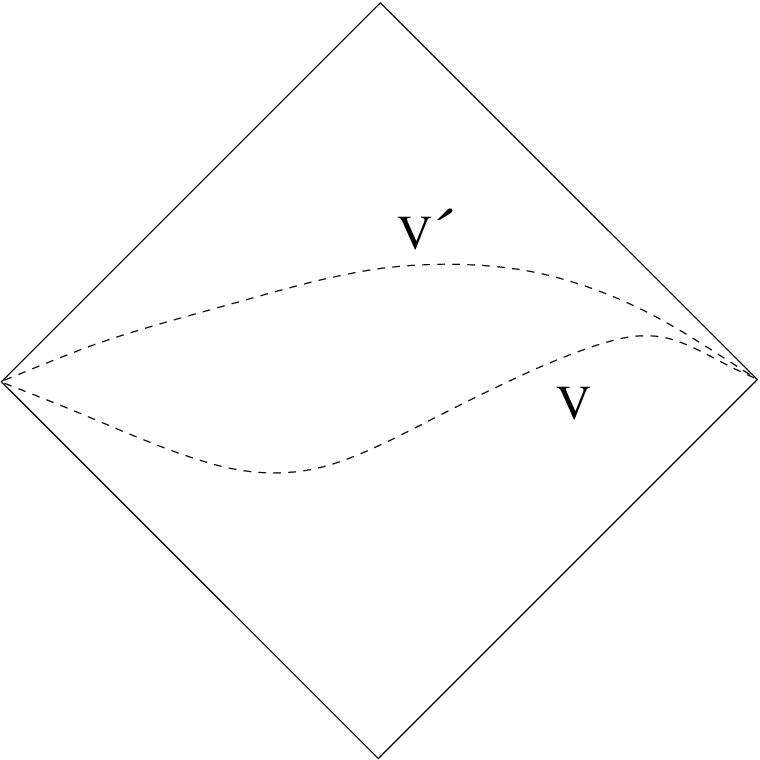}
\caption{Two spatial sets $V$ and $V^\prime$ in two dimensional Minkowski space. They have the same causal domain of dependence which is the diamond shaped set in the figure. Light rays are shown at $\pm 45^\circ$.}
\end{figure}

Write the two point correlation function restricted to $V$ as
\begin{equation}
C_{ij}(s_1,s_2)=\left\langle0 \right|\Psi_i(s_1)\,\Psi_j(s_2)^\dagger \left|0\right\rangle\,.
\end{equation}
 Then we must have
\begin{equation}
\textrm{tr} \left(\rho_V \Psi_i(s_1)\,\Psi_j^\dagger(s_2)\right)=C_{ij}(s_1,s_2).
\label{define}
\end{equation}
Taking into account the expressions (\ref{rho1}) and (\ref{rho2}) for the density matrix and the field algebra (\ref{algebra}) on the surface, the equation (\ref{define}) leads to the following relation between the  Hamiltonian kernel $H$ and the correlator
\begin{equation}
H(s_1,s_2)=-\bar{\eta}(s_1)\,\log\left[(C\, \bar{\eta})^{-1}-1)\right](s_1,s_2)\,.\label{gh}
\end{equation}
One can also summarize this in a more compact manner,
  \begin{equation}
 \rho_V=c\, e^{-{\cal H}}\,, \hspace{1.2cm} {\cal H}=\int_V ds_1\, ds_2\, \tilde{\Psi} (s_1) \tilde{H}(s_1,s_2) \Psi (s_2) \label{uno}\,,  
  \end{equation}
with 
\begin{equation}
\tilde{H}=-\log (\tilde{C}^{-1}-1 )
\end{equation}
and
\begin{equation}
\tilde{C}(s_1,s_2)=\left\langle0 \right|\Psi(s_1)\,\tilde{\Psi}(s_2) \left|0\right\rangle\,.\label{tree}
\end{equation}  
  
\subsection{Causality}
The explicit form of the density matrix in terms of the operators lying on any spatial curve allows us to show that the reduced density matrix does not change with the Cauchy surfaces having the same domain of dependence (see figure 1). Heuristically, this means the information cannot travel faster than the light speed, and in consequence it must be the same in the surfaces $V$ and $V^\prime$ in figure 1. In order to show this, consider the equation
\begin{equation}
\Psi_i(x)=\int_V ds \left\{\Psi_i(x),\tilde{\Psi}_j(s)\right\} \Psi_j(s)\,,
\end{equation}    
which is valid for any space-time point $x$ in the domain of dependence of $V$, and where $s$ is again a distance parameter along the curve $V$. This follows by the uniqueness of the solution of the Dirac equation since the right hand side satisfies the Dirac equation in $x$, and on the surface $V$ the equation is satisfied due to (\ref{algebra}). Then, let us call
\begin{equation}
G_{i,j}^{V^\prime V}(s_1,s_2)=\left\{\Psi_i(s_1),\tilde{\Psi}_j(s_2)\right\}\,,
\end{equation}
where $\Psi_i(s_1)$ is a field on $V^\prime$ and $\tilde{\Psi}_j(s_2)$ is a field on $V$. We have the following identities in operator notation
\begin{eqnarray}
\Psi^{V^\prime}&=&G^{V^\prime V} \, \Psi^V \,,\label{un}\\
\tilde{\Psi}^{V^\prime}&=&\tilde{\Psi}^V \, \tilde{G}^{V V^\prime} \,,\label{do}\\
\tilde{C}^{V^\prime V^\prime}&=& G^{V^\prime V} \tilde{C}^{V V} \tilde{G}^{V V^\prime}\,,\label{tre}
\end{eqnarray}
where
\begin{equation}
\tilde{G}^{V V^\prime}= \bar{\eta}^{V} (G^{V^\prime V})^\dagger \bar{\eta}^{V^\prime}\,.
\end{equation}
Now, considering the anticommutator in the surface $V^\prime$
\begin{equation}
{\bf 1}^{V^\prime}=\left\{\Psi^{V^\prime} , \tilde{\Psi}^{V^\prime}\right\}=G^{V^\prime V} \left\{\Psi^{V} , \tilde{\Psi}^{V}\right\}\tilde{G}^{V V^\prime}=G^{V^\prime V} \tilde{G}^{V V^\prime}\,.
\end{equation}
Thus, $G^{V^\prime V}$ and  $\tilde{G}^{V V^\prime}$ are inverse to each other. It is then easy to see that the similarity transformation of the correlator (\ref{un}-\ref{tre}) cannot change ${\cal H}$ in eqs. (\ref{uno}-\ref{tree}). This means that we can use any spatial slice for a 
given causally complete set (that is, a diamond shaped set arising as a domain of dependence of a spatial set) in order to compute the density matrix. 

\section{The massless case}  

In this paper we focus on a Dirac field in $1+1$ dimensions. The correlator for two points $x$ and $y$ with spatial separation $(x-y)^2< 0$ is 
\begin{equation}
C(x,y)=\frac{1}{4} \partial_\mu \left[\theta{(x-y)^2}\textrm{sgn}(x^0-y^0)\right] \, \gamma^\mu\gamma^0  +\frac{m}{2\pi} K_0(m|x-y|) \gamma^0  -\frac{i m}{2 \pi} \, K_1(m |x-y|) \frac{(x-y)_\mu}{|x-y|}\gamma^\mu \gamma^0 \, \,,\label{yuyi}
\end{equation}
where $K_a(x)$ is the standard modified Bessel function. The first term gives just  $1/2\, \delta(s_1-s_2)$ in $\tilde{C}$ for any surface.

In the massless case the problem can be decomposed into the different chiralities. Let us define the chirality operator $\gamma^3=\gamma^0 \gamma^1$, with $\gamma^3=(\gamma^3)^\dagger$ and $(\gamma^3)^2=1$, and the chirality projectors $Q_\pm=(1\pm \gamma^3)/2$. Let us also introduce the null coordinates
\begin{equation}
u_\pm=t\pm x\,.
\end{equation}
The massless Dirac equation leads to
$\partial_{u_\pm} \Psi_\mp =0$,
where $\Psi_\pm=Q_\mp \Psi$ are the field components of definite chirality (note that $\Psi_\pm$ has chirality $\mp 1$). Then we have  
\begin{equation}
\Psi=\Psi_+(u_+)+\Psi_-(u_-)\,.
\end{equation} 
These fields depend only on one of the null coordinates, and satisfy the anticommutation relations
\begin{eqnarray}
\left\{\Psi_\pm(u^1_\pm),\Psi_\pm^{\dagger}(u^2_\pm)\right\}&=&\delta(u^1_\pm-u^2_\pm)\,,\label{also1}\\
\left\{\Psi_\pm(u^1_\pm),\Psi_\mp^{\dagger}(u^2_\mp)\right\}&=&0\,.\label{also2}
\end{eqnarray}

Using this decomposition and the massless limit of the correlator (\ref{yuyi}) in eqs.(\ref{uno}-\ref{tree}) we have (for any choice of surface)
\begin{eqnarray}
\rho &=& \rho(u_+) \otimes \rho(u_-)=c \,\,e^{-{\cal H}_{+} }\otimes e^{-{\cal H}_{-} }\,, \label{ro}\\
{\cal H}&=&{\cal H}_{+} + {\cal H}_{-} \,, \\
{\cal H}_{\pm}&=& \int_{V_{\pm}} du_{\pm}^1 \, du^2_{\pm}\, \Psi_\pm^\dagger(u^1_\pm) H_\pm (u^1_\pm,u^2_\pm) \Psi_\pm(u^2_\pm)\,,\\
H_\pm&=& -\log (D_\pm^{-1}-1 )\,.\label{rot}
\end{eqnarray}
Here $V_{\pm}$ are the projections of $V$ in the $u_+$ and $u_-$ coordinate axis respectively, and $D_\pm$ is now the scalar kernel
\begin{equation}
D(x,y)=\frac{1}{2} \delta(x-y) - \frac{i}{2 \pi} \,  \frac{1}{(x-y)}\,\label{tuyo}
\end{equation}
where the domain of the real variables $x, y$ is $V_{\pm}$. Here and in the following the distribution ${(x-y)}^{-1}$ is taken with the principal value regularization. 

\subsection{Eigenvectors and the resolvent of $D$}
Therefore, in the massless case, the problem of finding an explicit expression for the local density matrix and related quantities is equivalent to the one of solving the operator with kernel $D$.  
   Fortunately, the resolvent of the kernel $(x-y)^{-1}$ in multicomponent subsets of the real line is known from the theory of singular integral equations of the Cauchy type \cite{singu}. We write in a generic way a set $V=(a_1 , b_1 )\cup (a_2 ,b_2 ) \cup ... \cup (a_n , b_n)$, formed by $n$ disjoint intervals, where $a_i< b_i$, $b_i< a_{i+1}$. For a set of intervals $I_i$ in spacetime, ordered from left to right in the spatial coordinate, we write the $(u_+,u_-)$ coordinates of the left and right extreme points of the intervals as $(u_{+i}^{L},u_{-i}^{L})$ and $(u_{+i}^{R},u_{-i}^{R})$. Thus in the following $(a_i, b_i)$ may mean either $(a_i^+,b_i^+)=(u_{+i}^{L}, u_{+i}^{R})$ or $(a_i^-,b_i^-)=(u_{-(n+1-i)}^{R}, u_{-(n+1-i)}^{L})$, according to where we are referring to the $u_+$ or the $u_-$ coordinate axis (note that in this later the order of the points is inverted). In this section we use generically the variables $x$, $y$ as meaning either coordinates on the null axis $u_+$ or in the axis $u_-$. 
   
 The resolvent of $D$,
   \begin{equation}
   R^0(\beta)=(D-1/2+\beta)^{-1}\label{nu}\equiv\left( - \frac{i}{2\pi}\frac{1}{x-y} +\beta \,\delta(x-y)\right)^{-1}\,,
   \end{equation}
 has the following expression  \cite{singu}
\begin{equation}       
R^0(\beta) =\left(\beta^2-1/4 \right)^{-1}
\left(\beta\,\delta(x-y)\, 
+\frac{i }{2\pi}   \frac{e^{-\frac{i}{2\pi} \log\left(\frac{\beta-1/2}{\beta+1/2}\right)\, (z(x)-z(y)) }}{x-y}
\right)\,,\label{reso}
\end{equation}
where
\begin{equation}
z(x)=\log\left(-\frac{\prod_{i=1}^n (x-a_i)}{\prod_{i=1}^n (x-b_i)}\right) \,. \label{bfbf}
\end{equation}

Moreover, we have been able to obtain the complete spectrum of $(x-y)^{-1}$ in $V$. We have guessed the solutions for the eigenvectors (somehow inspired in some formulae of the book \cite{singu}, and also using numerical integration) and checked directly they where the right eigenvectors, by integration. 

It has $n$ degenerate eigenvectors 
\begin{equation}
\int_V dy \, \frac{1}{x-y} \Psi^k_s(y)=\lambda_s \Psi^k_s(y)\,,
\end{equation}
$k=1,..,n$, corresponding to the eigenvalue 
\begin{equation}
\lambda_s =i \pi \tanh(\pi s) \label{eigen}
\end{equation}
for any $s\in R$. The eigenvalue corresponding to $H$ is $2\pi s$. Eq. (\ref{eigen}) and most of the identities below can be proved by  writing the integrals in the complex $z$ plane.
 
The eigenvectors can be chosen to form an orthonormal basis as 
\begin{equation}
\Psi_s^k(x)=\frac{\Theta_V(x) \prod_{i\neq k} (x-a_i)}{N_k\, \left(\prod_{i=1}^n |x-a_i| |x-b_i|\right)^{1/2}}\, e^{-i\, s\, z(x)}\,,\label{hfhf}
\end{equation}
where $\Theta_V(x)$ is a function which takes the value $(-1)^{j+1}$ in the $j^{\textrm{th}}$ interval $(a_j,b_j)$.
The normalization factor is
\begin{equation}
N_k=\sqrt{\pi} \left(\sum_{j=1}^n  \frac{\prod_{l\neq k} (b_j-a_l)}{(b_j-a_k) \prod_{l\neq j} (b_j-b_l)}-\frac{\prod_{j\neq k} (a_k-a_j)}{\prod_{j=1}^n (a_k-b_j)}\right)^{\frac{1}{2}}\,.
\end{equation}
We have from this
\begin{equation}
\sum_{i=1}^n N_i^{-2}=\frac{\sum_{i=1}^n (b_i-a_i)}{2\pi}=\frac{L_t}{2\pi}\,,\label{nuno}
\end{equation}
where $L_t$ is the sum of the $n$ interval lengths.
The vectors (\ref{hfhf}) then satisfy
\begin{equation}
\int_V dx \, \Psi^{k*}_s(x) \Psi^{k^\prime}_{s^\prime}(x)=\delta_{k,k^\prime} \, \delta(s-s^{\prime})\,, 
\end{equation}
and form a complete orthonormal basis
\begin{equation}
\sum_{k=1}^n \int_{-\infty}^\infty ds\, \Psi^{k*}_s(x) \Psi^{k}_{s}(y)=\delta(x-y)\,.
\end{equation}

We also have the useful relations
\begin{equation}
\int_V dx\, \Psi^{k}_s(x)=(-1)^{n+1} \pi \, \textrm{sech}(\pi s) N_k^{-1}\,,\label{nos}
\end{equation}
and 
\begin{equation}
\textrm{Im}\int_V dx\,x\, \Psi^{k}_s(x)= (-1)^{n}  s \pi \,\textrm{sech}(\pi s) N_k^{-1}    L_t\,.\label{uya}
\end{equation}

\subsection{Modular Hamiltonian}

The Hamiltonian kernel $H=-\log (D^{-1}-1 )$ writes in terms of the resolvent (\ref{nu}) as
\begin{equation}
H=-\int^\infty_{1/2} d\beta \, \left(R^0(\beta)+R^0(-\beta)\right)\,.
\end{equation}
Changing the variable $\beta$ to $\sigma=\frac{1}{2\pi} \log(\frac{\beta-1/2}{\beta+1/2})$, we have from (\ref{reso})
\begin{equation}
H=-i \int^\infty_{-\infty} d\sigma \, \frac{1}{x-y} \, e^{-i \sigma (z(x)-z(y))} \label{ss}=-2\pi i \,\frac{\delta\left(z(x)-z(y)\right)}{x-y}\,,
\end{equation}

The expression in (\ref{ss}) has two different contributions,
\begin{equation}
H=H_{\textrm{loc}}+H_{\textrm{noloc}}\,.
\end{equation}
$H_{\textrm{noloc}}$ is non local, and occurs when $z(x)=z(y)$ but $x\neq y$. The local one $H_{\textrm{loc}}$ is due to the solution $x=y$ of the equation $z(x)=z(y)$. After changing the variables in the argument of the $\delta$ function, this last contribution gives a product of distributions of the form $\delta(x-y)/(x-y)$. The different regularizations of this quantity must vanish outside $x=y$, and give $\delta(x-y)$ when multiplied by the function $(x-y)$.  The only solutions are of the form  $-\delta^\prime(x-y)+k\, \delta(x-y)$, where $k$ is regularization dependent. Here, $k$ is fixed by requiring that $H$ is hermitian. This gives for the local term
\begin{equation}
H_{\textrm{loc}}=\pi i \left( 2 \left(\frac{d z(x)}{dx}\right)^{-1}  \, \partial_x 
+ \frac{d}{dx}\left(\frac{d z(x)}{dx}\right)^{-1}\right) \,\delta(x-y)\,.
\end{equation}

The non local term follows by solving the equation $z(x)=z(y)$ for $x\neq y$. The function $z(x)$ increases monotonically inside each of the intervals $I_i$, being $-\infty$ on the points $a_i$ and tending to $\infty$ on the other end point  $b_i$ of the interval. Thus, it follows that given any $z\in(-\infty,\infty)$, there is exactly one solution $x_l(z)\in I_l$ of (\ref{bfbf}) for each interval. These solutions are given by the roots of a polynomial. Note that the different points $x_l(z)$ which are related to each other tend to the left (right) of the respective intervals simultaneously.  
In the case of two intervals the points connected to each other in the Hamiltonian are
\begin{equation}
x_1(z)=\frac{(b_1 b_2-a_1 a_2)\, x_2(z) +(b_1+b_2) a_1 a_2 - (a_1+a_2) b_1 b_2}{(b_1+b_2-a_1-a_2)\, x_2(z) +a_1 a_2 -b_1 b_2}\,.
\end{equation}
This is a global conformal transformation of square $1$, sending $a_1$ to $a_2$ and $b_1$ to $b_2$. 

The non local part of the Hamiltonian then writes
\begin{equation}
H_{\textrm{noloc}}=-2\pi i  \sum_{l, \,x_l(z(x))\neq x} \frac{1}{(x-y)} \, \left(\frac{dz}{dy}\right)^{-1} \delta(y-x_l(z(x)))\,.
\end{equation}
Remarkably, this non local Hamiltonian only mixes a finite number of points, the $x_l(z)$, one from each interval, having the same $z$.  

It is possible to obtain the same expressions for $H$ using directly the decomposition in terms of the eigenvectors
\begin{equation}
H(x,y)=2 \pi \sum_k\int_{-\infty}^{\infty} ds\, s\, \Psi^k_s(y) \Psi^{k*}_s(x)\,.
\end{equation}

\begin{figure} [tbp]
\centering
\leavevmode
\epsfysize=3.5cm
\bigskip
\epsfbox{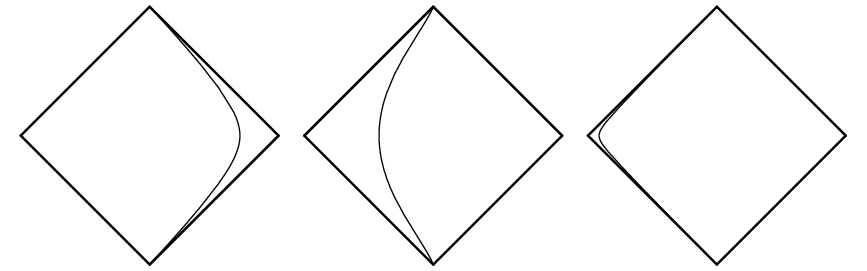}
\caption{Three related trajectories for a three interval case. The null coordinates in the three diamonds are $u_\pm^l(z_\pm)$, $l=1$, $2$, $3$, and $z_\pm\in (-\infty,\infty)$, with $z_+-z_-=$constant. The related points do not show any geometric symmetry in general.}
\end{figure}

\subsection{Internal dynamics}
The internal dynamics inside $V$ (one parameter group of modular transformations) is given by the unitary operators $\rho^{i \tau}$. In order to see how these transformations act we look at their action on the field operators
\begin{equation}
\Psi(x,\tau)=\rho^{i \tau}\Psi(x)\rho^{-i \tau}\,.
\end{equation}
Using the expressions (\ref{ro}-\ref{rot}) of the density matrix in terms of a quadratic Hamiltonian and the anticommutation relations (\ref{also1}-\ref{also2}), we have that the chiral components $\Psi_\pm(u_\pm,\tau)$ satisfy a Dirac-type equation with Hamiltonian $H_\pm$,
\begin{equation}
\frac{d\Psi_\pm}{d\tau}=-i \left[{\cal H_\pm},\Psi_\pm  \right]=i H_\pm \Psi_\pm\,.\label{dir}
\end{equation}

For one interval it is known that $H=H_{\textrm{loc}}$ is proportional to the generator of global conformal transformations which keep the interval fixed. Eq. (\ref{dir}) leads in this case to the point transformation for the operators  \cite{haag,confor,uninterval}   
\begin{eqnarray}
\Psi_\pm(u_\pm(\tau),\tau )&=&\sqrt{\frac{(u_\pm^0-a_\pm)(b_\pm-u_\pm^0)}{(u_\pm(\tau)-a_\pm)(b_\pm -u_\pm(\tau))}}\,\,\Psi_\pm(u_\pm^0)\,,\\
u_\pm(\tau)&=&\,\frac{ b_\pm(u_\pm^0-a_\pm)+e^{-2\pi \tau} a_\pm(b_\pm-u^0_\pm)}{ (u^0_\pm-a_\pm)+e^{-2\pi \tau} (b_\pm-u^0_\pm)}\,, \label{fl}
\end{eqnarray}
where $(a_\pm,b_\pm)$ are the projections of the interval in the $u_\pm$ coordinate axis ($a_+$ and $b_-$ correspond to the left endpoint while $a_-$ and $b_+$ correspond to the right one). 

For more than one interval the solution of (\ref{dir}) can be given by expanding in eigenvectors of $H_\pm$. However, these expressions are rather obscure. In order to understand the local dynamics let us write the field in the related points in the different intervals as a vector of fields, $\Psi_l(z)\equiv \Psi(x_l(z))$. Then eq. (\ref{dir}) writes (the same equation applies for both chiralities, we omit the $\pm$ subindices for notational convenience)
\begin{equation}
\frac{\partial \Psi_l}{\partial \tau}=-2\pi \frac{\partial \Psi_l}{\partial z}-  \pi  \frac{d}{dz}\left(\log(\frac{dx_l}{dz})\right) \Psi_l+ 2\pi \sum_{l^\prime \neq l} \frac{dx_{l^\prime}}{dz} \frac{\Psi_{l^\prime}}{x_l-x_{l^\prime}}\,.\label{ytyt}
\end{equation}
The characteristic trajectories $z(\tau)$ of the first order differential equation are 
\begin{equation}
z(\tau)=z_0+2 \pi \tau\,, \label{tray}
\end{equation}
for a constant $z_0$. 
Thus, the fields $\Psi_l(\tau)\equiv \Psi_l(z(\tau),\tau)$,  
 will mix to each other for different $l$ along the characteristic. 
Only the $n$ trajectories $x_l(z(\tau))$, with a given fixed $z_0$ will mix to each other, and the movement is such that $z$ moves at constant speed $2\pi$ with respect to $\tau$. 
 The figure (2) shows a group of three related trajectories inside the domain of dependence of three intervals (the three diamond shaped sets). In figure (3) the flux of trajectories are shown for two diamonds in three different relative positions.

\begin{figure} [t]
\centering
\leavevmode
\epsfysize=10cm
\bigskip
\epsfbox{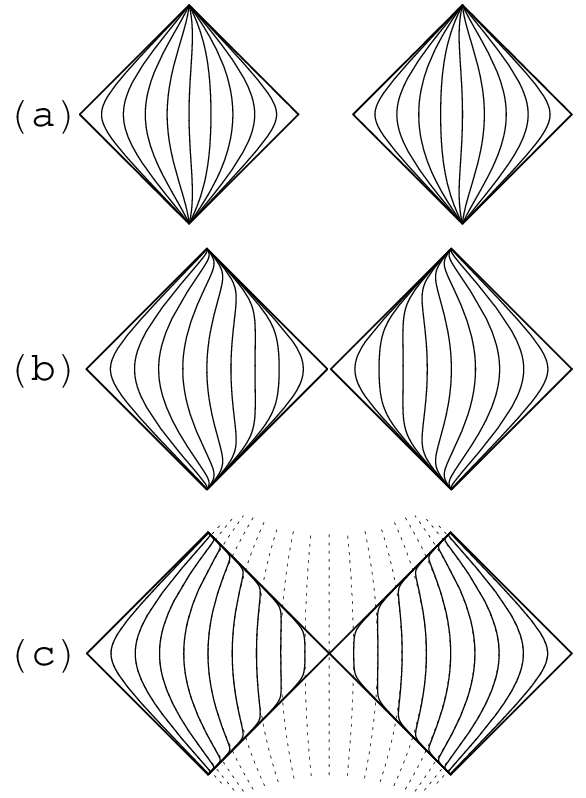}
\caption{Two intervals of size $d$ separated by: (a) $(1/4) d$  (top pair of diamonds), (b) $(1/70) d$, and (c) $(1/1000) d$ (bottom pair of diamonds). The trajectories in (a) approach the ones corresponding to single independent diamonds. In (b) the curves get distorted by the proximity of the diamonds. In (c) the curves are very similar to the ones of a single diamond of size $2 d$ (shown with dashed lines), excepting when they hit the null boundaries and run along them afterward. The trajectories shown for the two diamonds are equal spaced in the $t=0$ plane, and they are not related to each other.}
\end{figure} 
  
On a group of related trajectories the $\Psi_l(\tau)$ evolve with $\tau$ according to a coupled set of ordinary differential equations. Writing 
\begin{equation}
\tilde{\Psi}_l(\tau)\equiv (d x_l/dz)^{1/2}\Psi_l(\tau)\,,\label{kambaquerido}
\end{equation}
 we have
\begin{equation}
\frac{d\tilde{\Psi}_l}{d\tau} =2\pi\sum_{l\neq l^\prime} \frac{\left(\frac{dx_l}{dz}\right)^{\frac{1}{2}}\left(\frac{dx_{l^\prime}}{dz}\right)^{\frac{1}{2}}}{x_l-x_{l^\prime}}\tilde{\Psi}_{l^\prime}\,,\label{baile}
\end{equation}
where $x_l$ and $z$ are functions of $\tau$ according to (\ref{tray}). Note that since the matrix on the right hand side is antisymmetric, the sum $\sum_l (\tilde{\Psi}_l(\tau))^2$ is a constant, and $\tilde{\Psi}(\tau)={\cal O}(\tau) \tilde{\Psi}(0)$, with ${\cal O} (\tau)$ orthonormal.

The analytic solution of this equation for $n=2$ can be obtained explicitly.  In this case we write 
\begin{equation}
 {\cal O}=\left(\begin{array}{cc}
  \cos(\theta) & -\sin(\theta)\\
  \sin(\theta) & \cos (\theta)
  \end{array}\right)\,,
  \end{equation}
and the solution of the differential equation (\ref{baile}) gives\footnote{We thank the authors of \cite{Longo} for pointing out an error in the previous version of this formula. }
  \begin{equation}
   \theta(\tau) =\arctan \left[ \frac{x_1(\tau) (b_1+b_2-a_1-a_2)}{\sqrt{(b_1-a_1)(a_2-b_1)(b_2-a_1)(b_2-a_2)}} \right]-\arctan \left[ \frac{x^0_1 (b_1+b_2-a_1-a_2)}{\sqrt{(b_1-a_1)(a_2-b_1)(b_2-a_1)(b_2-a_2)}} \right].
  \end{equation}
The absolute value of the angle $\theta$ never reaches $\pi/2$, meaning that the rotation of components never completes, there is no complete "teleportation" or swap between the components. However, $\pi/2$ is approached as the intervals come close to each other. 

\section{Entropy}
From (\ref{eigen}) it follows that the spectrum of $\rho_V$ is continuous and covers all the interval $(0,1)$. This means that even if $\rho_V$ is well defined as a state on the operator algebra, it is not a bona-fide density matrix, since it is not trace class.  Then, the expression for the entropy which follows from (\ref{uno}-\ref{tree}),
\begin{equation}
S(V)=-\textrm{tr}\left[(1-\tilde{C})\log(1-\tilde{C})+\tilde{C}\log(\tilde{C})\right]\,,
\end{equation}
has to be regularized. We will find more convenient to express the entropy in terms of the resolvent $R=(\tilde{C}-1/2+\beta)^{-1}$ as
\begin{equation}
S(V)=-\int^\infty_{1/2} d\beta\, \textrm{tr}\left[\left(\beta-1/2\right) \left(R(\beta)-R(-\beta)\right)-\frac{2\beta}{\beta+1/2}\right]\,.\label{fas}
\end{equation}  
With the knowledge of the spectrum and the resolvent for the massless case we are in position to compute the entropy and to do some expansions for the massive case.

\subsection{Massless case}
In this case we use directly the chiral decomposition and the resolvent (\ref{reso}). In applying (\ref{fas}) the term proportional to the identity cancels with the corresponding one in the resolvent. Then we have, for any chiral component, a contribution of the form
\begin{equation}
S(V)=-\frac{1}{\pi}\int^\infty_{1/2} d\beta\, \int_V dx\, \lim_{y\rightarrow x}  \frac{\sin\left[ \frac{1}{2\pi} \log\left(\frac{\beta-1/2}{\beta+1/2}\right)\, (z(x)-z(y)) \right]}{(\beta+1/2)\,(x-y)}\,,
\end{equation}
where we are using the notations of sections 3.1 and 3.2, and here $x$, $y$ should be understood as coordinates on one of the chiral axis.     
 One can integrate first over $\beta$, giving
\begin{eqnarray}
S(V)&=&\int_V dx\, \lim_{y\rightarrow x}  \frac{\frac{z(x)-z(y)}{2}\coth((z(x)-z(y))/2)-1}{(x-y)(z(x)-z(y))}=\frac{1}{12}\int_V dx\,\sum_{i=1}^n \left(\frac{1}{x-a_i}-\frac{1}{x-b_i}\right)\nonumber\\
&\,&\hspace{1cm}=\frac{1}{6} \left( \sum_{i,j}\log|b_i-a_i|-\sum_{i<j} \log|a_i-a_j| -\sum_{i<j} \log|b_i-b_j|-n \log \epsilon \right)\,.\label{sesen}
\end{eqnarray}
Here $\epsilon$ is a distance cutoff introduced in the last integration, which extends from $a_i+\epsilon$ to $b_i-\epsilon$ in each interval. The divergent terms on the entanglement entropy are associated to the boundary of the region, where the high frequency modes give place to an unbounded entanglement with the outside region. Physically sensible cutoff independent quantities can be obtained from the regularized entropy by appropriate subtractions. For example, the mutual information $I(A,B)=S(A)+S(B)-S(A\cup B)$ is finite and well defined, since the boundary terms cancel in this combination.

This same result was obtained by bosonization in \cite{fermion}. The formula (\ref{sesen}) applies for each chirality. The same formula but with coefficient $1/3$ instead of $1/6$ gives the entropy for a set $V$ on a spatial line (two chiralities).

\subsection{Massive case: small mass limit}
 For the massive case we consider only sets lying in the same spacial line, and thus the correlator reads 
\begin{equation}
C(x,y)=\frac{1}{2} \delta(x-y) +\frac{m}{2\pi} K_0(m|x-y|) \gamma^0  +\frac{i m}{2 \pi} \, K_1(m (x-y)) \gamma^3  \,.\label{crosta}
\end{equation}

The short distance expansion of the correlator up to second order in the field mass is $C=C_0+C_1+C_2+...$ where
\begin{eqnarray}
C_0 (x,y)&=&\frac{1}{2} \delta(x-y)\, 
 \mathbb I+ \frac{i}{2\pi}\frac{1}{x-y} \gamma^3\,, \label{horacio}\\
C_1 (x,y)&=&-\frac{m}{2\pi}\left(\gamma_E+\log\left(\frac{m|x-y|}{2}\right)\right) \gamma^0 \,,\\
C_2 (x,y)&=&\frac{i m^2}{4\pi} (x-y) \left(\gamma_E -\frac{1}{2} +\log\left(\frac{m|x-y|}{2}\right) \right) \gamma^3\,,
\end{eqnarray}
and $\gamma_E$ is the Euler constant.
The perturbative expansion of the resolvent for small mass is done by considering the formula
\begin{equation}
R_V(\beta)=R_V^0(\beta)-R_V^0(\beta)C_1 R_V^0(\beta)-R_V^0(\beta)C_2 R_V^0(\beta)+R_V^0(\beta)C_1 R_V^0(\beta)C_1 R_V^0(\beta)-...\,.
\label{financiar}
\end{equation}
The first term gives the massless contribution (\ref{sesen}). The second one is of first order in the mass, and   
 does not contribute to the entropy since it is traceless. From (\ref{horacio}-\ref{financiar}) the expansion of the entropy has the form
 \begin{equation}
 S=\sum_{i=0}^\infty \sum_{j=0}^{2 i} s_{i,j}\,\, m^{2 i}\log^j(m)\,.
 \end{equation}

Let us calculate the contributions which are second order in the mass coming from the third and fourth terms. First we calculate the leading $\log^2(m)$ term. We find convenient to use the expression of the resolvent in terms of the eigenvectors
$\Psi_s^k(x)$ of $(x-y)^{-1}$. From (\ref{horacio}) we have
\begin{equation}
R_V^0(\beta)(x,y)=\sum_{k=1}^n\int^{+\infty}_{-\infty} ds\, \Psi_s^{k}(x) \, M(\beta , s)\,   \Psi_s^{k\,*}(y)\,,
\end{equation} 
where
\begin{equation}
M(\beta , s)=\left(\beta\,  \mathbb I -\frac{\tanh(\pi s)}{2}\,\gamma^3 \right)^{-1}\,.
\end{equation}

Thus, the $\log^2(m)$ contribution which comes from the fourth term in (\ref{financiar}) is
\begin{eqnarray}
\delta S_{\,2,2}&=&s_{2,2}\, m^2\log^2(m)= -\frac{m^2}{(2\pi)^2}  \log^2(m)\int^\infty_{1/2} d\beta\,\int^\infty_{-\infty} ds\, \int^\infty_{-\infty} ds^\prime\, (\beta-1/2) \nonumber \\
&& \sum_{k=1}^N \sum_{k^\prime=1}^N |\Lambda(s,k)|^2|\Lambda(s^\prime,k^\prime)|^2 \, \textrm{tr} (M(\beta , s) \gamma^0 M(\beta , s^\prime)^2 \gamma^0)\,,\label{arra}
\end{eqnarray}
where
\begin{equation}
\Lambda(s,k)=\int_V  dx\,  \Psi_s^k(x)\,.
\end{equation}
From (\ref{nuno}) and (\ref{nos}) we have 
\begin{equation}
\sum_{k=1}^N |\Lambda(s,k)|^2=\frac{\pi}{2} \textrm{sech}^2(\pi s) L_t\,.
\end{equation}
Using this, the integrals in (\ref{arra}) can be found exactly giving
\begin{equation}
\delta S_{\,2,2}=-\frac{m^2 L_t^2}{6}  \log^2(m)\,.\label{leading}
\end{equation}

The next contributions are of order $m^2 \log (m)$. They come from the third and fourth terms in (\ref{financiar}). We can divide this contribution in three terms,
\begin{equation}
\delta S_{\,2,1}=s_{2,1}\,\,m^2 \log(m)=\Delta_1+\Delta_2+\Delta_3\,.
\end{equation}
 The contribution coming from the terms in $C_1$ which does not contain $\log|x-y|$ we call $\Delta_1$, the one from $C_2$, we call $\Delta_2$, and the last one $\Delta_3$ which involves  $\log|x-y|$ in $C_1$. $\Delta_1$ is readily evaluated since the relevant calculation is the same as above,
\begin{equation}
\Delta_1=\frac{\log(2)-\gamma_E}{3}\, \,m^2 L_t^2 \,\,\log(m)\,.
\end{equation}  
The contribution of $C_2$ to this order is
\begin{eqnarray}
\Delta_2&=& \frac{i m^2}{2\pi}  \log(m) \int^\infty_{1/2} d\beta\,\int^\infty_{-\infty} ds\,  (\beta-1/2)  \, \textrm{tr} (\gamma^3\,M(\beta , s)^2)\nonumber \\
&&\hspace{5.3cm}\times \sum_{k=1}^N \int_V dx\,\int_V dy\, (x-y) \Psi_s^{k\,*}(x) \Psi_s^k(y)\,.\label{arra1}
\end{eqnarray}
Using (\ref{nuno}), (\ref{nos}) and (\ref{uya}) we find
\begin{equation}
\Delta_2=-\frac{1}{6} \, \,L_{\textrm{t}}^2 \,m^2\,\log(m)\,.
\end{equation}

Finally, the expression for $\Delta_3$ is
\begin{eqnarray}
\Delta_3= -\frac{m^2}{4\pi^2}  \log(m)\int^\infty_{1/2} d\beta\,\int^\infty_{-\infty} ds\, \int^\infty_{-\infty} ds^\prime\, \sum_{k=1}^N \sum_{k^\prime=1}^N(\beta-1/2) \Lambda(s,k)\Lambda(s^\prime,k^\prime)^* \nonumber \\
  \, \left(\textrm{tr} (M(\beta , s)^2 \gamma^0 M(\beta , s^\prime) \gamma^0)-(\beta\leftrightarrow -\beta)\right) \int_V dx\,\int_V dy\,  \Psi_s^{k\,*}(x) \log|x-y| \Psi_{s^\prime}^{k^\prime}(y)+\textrm{hc} \,.\label{arra2}
\end{eqnarray} 
We first do the integrals in $s$ and $s^\prime$. Then, the integral in $\beta$ gives a term antisymmetric in $x$ and $y$, which is discarded, and a single symmetric term proportional to the delta function $\delta(z(x)+z(y))$. Using the identity
\begin{equation}
\sum_{k=1}^n \frac{1}{N_k^2 (x-a_k)}=\frac{1}{2\pi} (1+e^{-z})\,,
\end{equation}
we find after some algebra
\begin{eqnarray}
\Delta_3&=& - 2 m^2  \log(m) \int_V dx\,\int_V dy\, \log|x-y|\, \delta(z(x)+z(y))\nonumber \\
&=& 2 m^2  \log(m) \int_{-\infty}^\infty dz\, \sum_{l,l^\prime=1}^N \log|x_l-\tilde{x}_{l^\prime}| \, \frac{dx_l}{dz} \, \frac{d\tilde{x}_{l^\prime}}{dz}\,.
\end{eqnarray}
In the last equation we have written $x_l(z)$ as the only solution of $z(x)=z$ in the $l^{\textrm{th}}$ interval, and $\tilde{x}_l(z)$ as the only solution of $z(x)=-z$ in the $l^{\textrm{th}}$ interval. For a single interval of length $L$ this gives 
\begin{equation}
\Delta_3=-\frac{1}{3} \log(L) \, \log(m)\,  m^2 L^2+ \frac{4}{9} \log(m)\, m^2 L^2\,,
\end{equation}
but for more than one interval $\Delta_3$ does not depend simply on the total length $L_t=\sum (b_i-a_i)$. In the case of a single interval the entropy can be expressed in terms of a solution of a Painlev\'e ordinary differential equation \cite{fermion}. The expansion for small mass obtained here coincides in this case with the one (more easily) obtained with the help of this differential equation.

A remarkable feature of the massless fermion entropy (\ref{sesen}) is that the mutual information $I(A,B)=S(A)+S(B)-S(A\cup B)$ is extensive in this case, $I(A,B\cup C)=I(A,B)+I(A,C)$ \cite{fermion}. The leading correction (\ref{leading}) for small mass $\sim m^2 \log (m)^2$ also has extensive mutual information. 
 The first non extensive contribution is $\Delta_3$ at order $m^2 \log(m)$ \cite{rema}.

\subsection{Massive case: long separation distance}
The case of two sets $A$ and $B$ which are separated by a distance $r$ large with respect to their sizes can be treated as in \cite{rema}.  Following \cite{rema} the resolvent can be expanded perturbatively,
\begin{equation}
R_V^r(\beta)=R_V^\infty(\beta)-R_V^\infty(\beta){\cal C}_1 R_V^\infty(\beta)+R_V^\infty(\beta){\cal C}_1 R_V^\infty(\beta){\cal C}_1 R_V^\infty(\beta)-...\,.
\label{finan1}
\end{equation}
Here $R_V^\infty(\beta)$ is the unperturbed resolvent, 
\begin{equation}
R_V^\infty(\beta)=\left(\begin{array}{cc}
R_A(\beta) & 0 \\
0 & R_B(\beta) 
 \end{array}
\right)\,, \label{tyto}
\end{equation}
corresponding to infinite separation distance, and ${\cal C}_1$ is the field correlator between points in $A$ and $B$
\begin{equation}
{\cal C}_1= \left(\begin{array}{cc}
0& C(\vec{r}) \\
 C^\dagger(\vec{r}) & 0 
 \end{array}
\right)\,.
\end{equation}

When $r$ is large $C(\vec{r})$ is small and the expansion (\ref{finan1}) makes sense. The correlator between points in one set and the other can be considered as a constant in the lowest order approximation (up to correction of the order of the cocient of the sises $L_{A,B}$ of $A$ and $B$ with $|\vec{r}|$). Thus, we have
\begin{equation}
{\cal C}_1= \left(\begin{array}{cc}
0& C(r)\, {\mathbf 1}_{A,B} \\
 C^\dagger(r) \,{\mathbf 1}_{B,A} & 0 
 \end{array}
\right)\,.
\end{equation}
The kernel ${\mathbf 1}_{A,B}(x,y)=1$ for any $x\in A$ and $y\in B$ and $C(r)$ is given by (\ref{crosta}) with $|x-y|=r$. 

 If $A$ and $B$ are small relative to the inverse mass, we can make use of the massless resolvent in (\ref{tyto}), as the zero order term of an expansion in powers of $(m \,L_{A,B})$ for $R_A(\beta)$ and $R_B(\beta)$. In this case we have for the mutual information
\begin{equation}
I(A,B)\sim \int^\infty_{1/2} d\beta\,\left(\beta-1/2\right)\left[\, \textrm{tr} \left( C^\dagger(r) \overline{R^0_A}(\beta) C(r) \overline{R_B^{0\,2}}(\beta) + (A\leftrightarrow B)\right)            -\left(\beta\rightarrow -\beta\right)\right]\,,\label{inn}
\end{equation}
where the bar over the resolvent and the square of the resolvent means sum over the spatial variables, $\overline{{\cal O}_X}=\int_X dx\, \int_X dy\, {\cal O}(x,y)$. Expanding the resolvents in terms of the eigenvectors as above, we arrive at the leading term 
\begin{equation}
I(A,B)\sim \frac{1}{3} \,m^2 \,L^A_t\, L^B_t \,(K_0^2(m r)+ K_1^2(m r)) \,\,\,\,(1 + {\cal O} (L_{A,B}/r)+{\cal O} (m^4 L_{A,B}^4)+...)\,.  \label{hghg}
\end{equation}
In this formula we have to use $L_{A,B}\ll r, m^{-1}$, but $mr$ can have any value.  Note that (\ref{hghg}) also shows an extensive mutual information at the first order in the long distance expansion. 
\section{Concluding remarks}
We have shown that the internal dynamics for a multicomponent region and a massless fermion in two dimensions is non-local. However, it displays what can be called a quasilocal behavior, since it only mixes the fields on a finite number of one dimensional trajectories. The $n$ related trajectories are determined by the parameter $z$ which moves at constant velocity. It is interesting to note that the relation $-e^z=\prod (x-a)/\prod (x-b)$ can be thought as a "uniformization" transformation \cite{cc}. This is the many to one transformation which is analytical except at the points $a_i$ or $b_i$, and maps each interval into the real line. Since this transformation is determined by analyticity properties alone, one could wonder if a similar phenomena of quasilocality might also occur for other conformal field theories in two dimensions. The factors (\ref{kambaquerido}) correspond to the standard general conformal transformation of the (primary) fermion operators. The picture is that of a single variable $z$ with uniform motion, and the $n$ trajectories generated by the multivalued transformation. The particularities of each theory might then be reflected in the dynamics of the field rotations between the different trajectories.     
\section*{Acknowledgments}
H.C. and M.H. thank CONICET, ANPCyT and UNCuyo for financial support.

\end{document}